\documentclass[prb,aps,amsmath,showpacs,preprint,]{revtex4}
\usepackage{graphicx}

\usepackage{dcolumn}

\usepackage{bm}
\usepackage{subfigure}
\usepackage{color}

\begin{document}

\title{Electronic topological transition in LaSn$_3$ under pressure}

\author{Swetarekha Ram$^1$, V. Kanchana$^{1,*}$,  G. Vaitheeswaran$^2$, A. Svane$^3$, S. B. Dugdale$^4$ and N. E. Christensen$^3$}

\affiliation{ $^1$ Department of Physics, Indian Institute of Technology Hyderabad, Ordnance Factory Estate, Yeddumailaram, Hyderabad-502 205, Andhra Pradesh, India. \\
$^2$ Advanced Centre of Research in High Energy Materials (ACRHEM), University of Hyderabad, Prof. C. R. Rao Road, Gachibowli, Hyderabad- 500 046,  Andhra Pradesh, India. \\
$^3$ Department of Physics and Astronomy, Aarhus University, DK-8000 Aarhus C, Denmark. \\
$^4$ H. H. Wills Physics Laboratory, University of Bristol, Tyndall Avenue, Bristol BS8 1TL, United Kingdom. \\
}
\date{\today}

\begin{abstract}
The electronic structure, Fermi surface and elastic properties of the iso-structural and iso-electronic LaSn$_3$ and YSn$_3$ intermetallic compounds are studied under pressure within
the framework of density functional theory including spin-orbit coupling. The LaSn$_3$ Fermi surface consists of two sheets, of which the second is very complex. Under pressure a third sheet
appears around compression $V/V_0=0.94$, while a small topology change in the second sheet is seen at compression  $V/V_0=0.90$. 
 This may be in accordance with the anomalous behaviour in the superconducting transition temperature observed in LaSn$_3$, which has been suggested to reflect a
 Fermi surface topological transition, along with a non-monotonic pressure dependence of the density of states at the Fermi level. 
 The same behavior is not observed in YSn$_3$, the Fermi surface of which already includes three sheets at ambient conditions, and the topology remains unchanged under pressure. 
The  reason for the difference in behaviour between LaSn$_3$ and YSn$_3$ is  the role of spin-orbit coupling and the  hybridization of La - $4f$ states with the Sn - $p$ states
in the vicinity of the Fermi level, which is well explained using the band structure calculation. 
The elastic constants and related mechanical properties are calculated at ambient as well as at elevated pressures. The elastic constants increase with  pressure for both  compounds and satisfy the conditions for mechanical stability under pressure.
\end{abstract}


\maketitle


\section{Introduction}
The RX$_3$-type ( R=rare earth elements, X= In, Sn, Tl, Pb) intermetallic compounds, which crystallize in the simple cubic Cu$_3$Au structure, have been
the subject of  many experimental and theoretical investigations because of their diverse properties. Many of these compounds
are superconductors. LaSn$_3$ and YSn$_3$ are particularly significant as they are found to have relatively high superconducting transition temperatures,
 T$_c$. For LaSn$_3$, T$_c=6.5$ K, \cite{Gambino,Bucher} and for YSn$_3$ T$_c= 7.0$ K, \cite{Kawashima} whereas LaPb$_3$, LaTl$_3$ and LaIn$_3$ have lower T$_c$'s of  4.05 K, 1.51 K and 0.71 K, respectively. \cite{Gambino,Bucher} 
Some of the RX$_3$ compounds, such as  PrSn$_3$
and NdSn$_3$, are found to order antiferromagnetically at  T$_N$= 8.6 K 
and 4.5 K, respectively,\cite{Shenoy} and CeSn$_3$ has been categorized as a
dense Kondo compound exhibiting valence fluctuations.\cite{Murani} 
It is interesting to compare the properties of LaSn$_3$ and YSn$_3$ to reveal to which extent the similar valence electron configurations of Y and La  influence the details of the electronic structure.
\cite{Borsa,Welsh,Toxen,Gambino,Havinga} 
The LaX$_3$ ( X= Sn, In, Tl, Pb) series and their alloys show an
oscillatory dependence in their bulk properties (superconducting transition temperature, magnetic susceptibility, thermoelectric power factor)  as
a function of average valence-electron number.\cite{Havinga,Havinga2,Grobman,Havinga3,Toxen2} 
At first, this behavior was interpreted in a nearly-free electron model as a reflection of the Fermi surface crossing the Brillouin zone close to the X-point,\cite{Havinga,Hoever} 
a viewpoint later contested by Grobman.\cite{Grobman} The  pressure dependence
of the critical temperature of LaSn$_3$ is anomalous, as shown by Huang et al.,\cite{HuangSSC} which these authors expect
to be driven by a Fermi surface topology change.

In the present study, we calculate the Fermi surface of LaSn$_3$ and indeed observe a change in topology under pressure, where a third set of electron pockets appear and
a minor part of the complex second sheet transfers from a closed orbit to an open orbit region. A similar transition is not predicted for  
 YSn$_3$, despite the overall similarity of their electronic structures. 
A number of
studies are available on the band structures of LaSn$_3$ and YSn$_3$,\cite{Hasegawa,Boulet,Yamagami,Dugdale} while less efforts have been devoted to the
pressure dependence  of the electronic structure, Fermi surface and elastic properties for these compounds. 
Hence we focus our attention in this paper on 
analysing the pressure induced
Fermi surface topology change in LaSn$_3$, which might be associated with the anomalous behaviour of T$_c$ under pressure,
 and present a comparative study
of LaSn$_3$ and YSn$_3$ under pressure. The remainder of the paper is organized as follows: Section II describes the method of
calculation used in this study. The results and discussions are presented in Section III, while section IV concludes the paper.

\section{Method of Calculation}
The calculations were performed  using the  full-potential linear augmented plane wave (FP-LAPW) method as implemented in the WIEN2K computer code,\cite{Blaha} based on density
functional theory (DFT), \cite{Hohenberg} which has been shown to yield reliable results for the electronic and structural properties of crystalline solids. 
Spin-orbit coupling (SOC) was included.
For the exchange-correlation functional, both the local density approximation (LDA) as proposed by Ceperley and Alder \cite{Ceperley} and generalized gradient approximation
(GGA) according to the Perdew-Burke-Ernzerhof \cite{PBE} parametrization
 was used. 
In order
to achieve energy eigenvalue convergence, the wave functions in the interstitial region were expanded using plane
waves with a cutoff of R$_{MT}$K$_{max}$= 9, where K$_{max}$ is the plane wave cut-off, and R$_{MT}$ is the smallest of all atomic sphere radii. The
charge density was Fourier expanded up to G$_{max}$=18 (a.u.)$^{-1}$. The maximum {$\ell$} value for the wave function expansion inside the atomic
spheres was confined to $\ell_{max}=10$.
Convergence tests were  carried out using higher G$_{max}$ and R$_{MT}$K$_{max}$ values, 
giving no significant changes in the calculated properties.
The muffin-tin radii were chosen as 2.75 a.u. for both La and Y and 2.83 a.u. for Sn. \textcolor{blue}{ \bf A ( 32 $\times$ 32 $\times$ 32 ) Monkhorst-pack\cite{Monkhorst} $k$-point mesh was used resulting in 396 
{\it k}-points in the irreducible part of the Brillouin zone.} 
The self-consistent calculations were considered to be converged when the total energy of the system was stable within 10$^{-6}$ Ry. The Birch-Murnaghan equation of states \cite{Birch}
was used to fit the total energy as a function of unit cell volume to obtain the equilibrium lattice
constants and bulk moduli for the investigated systems.
For the Fermi surfaces of RSn$_3$  a ( 64 $\times$ 64 $\times$ 64 ) mesh was used to ensure accurate determination of the Fermi level and smooth interpolation of the bands
crossing the Fermi level. The three dimensional (3D) Fermi surface plots were generated with the help of the Xcrysden molecular structure
visualization program. \cite{Kokalj}
\par

The elastic constants have been calculated within the total-energy method, where the unit cell is subjected to a number of small amplitude strains along
several directions. The elastic constants of solids provide links between the mechanical and dynamical properties of the crystals. In particular, they
provide information on the stability and stiffness of materials. It is well known that a cubic crystal has only three independent elastic constants
\cite{Nye,Mattesini,Soler}  C$_{11}$, C$_{12}$ and C$_{44}$.  From these one may obtain the Hill's \cite{Hill} shear modulus G$_H$, (which is the arithmetic mean of the Reuss\cite{Reuss} and Voigt\cite{Voigt} approximations), Young's modulus $E$, and the Poisson's ratio $\sigma$ by using stardard relations. \cite{kanchana}  
Furthermore, the Debye temperature may be obtained in terms of the mean sound velocity $v_m$:
\begin{equation}
\Theta_D =\frac{h}{k_B}\left(\frac{3n\rho N_A}{4\pi M}\right)^{1/3}v_m,
\end{equation}
 where $h$, $k_B$ and $N_A$ are Planck's, Boltzmann's constants,  and  Avogadro's number, respectively. $\rho$ is the mass density, $M$ the molecular weight, and
 $n$ the number of atoms in the unit cell.  The mean sound velocity  is defined as:
\begin{equation}
v_m = {\left[\frac{1}{3}\left(\frac{2}{v_t^3}+\frac{1}{v_l^3}\right)\right]}^{-1/3},
\end{equation}
where $v_l$ and $v_t$ are the longitudinal and transverse sound velocities, which may be  obtained from the shear modulus G$_H$ and  bulk modulus B as:
\begin{equation}
v_l =\sqrt{\frac{(B+\frac{4}{3}G_H)}{\rho}}
\end{equation}
and
\begin{equation}
v_t =\sqrt{\frac{G_H}{\rho}}.
\end{equation}

\section { Result and discussion }
\subsection {Ground state properties}
The RSn$_3$ compounds crystallize in the Cu$_3$Au type structure with space group $Pm\bar{3}m$ (No. 221). 
The calculated equilibrium lattice parameters $a$ and zero pressure bulk modulus $B$  are listed in Table I. 
The results are in good agreement with available experimental
 data and other calculations.  
 It often occurs that LDA underestimates and GGA overestimates the equillibrium lattice constants, which is also found in the present case, where the LDA value is $\sim 1$ \% below
 and the GGA value $\sim 1$ \% above the experimental value. 
 The somewhat lower bulk modulus obtained with GGA reflects the larger equilibrium volume in this approximation. 
 In the remainder of this study we report results on the basis of the LDA.

\subsection{ Elastic constants and mechanical properties }

The calculated elastic constants
(C$_{11}$, C$_{12}$ and C$_{44}$ ) at ambient pressure for LaSn$_3$ and YSn$_3$ are presented in Table II, together with quantities related to the elastic constants. 
The calculations were performed at
the equilibrium lattice constant as calculated with the LDA. The elastic constants extracted from the experimental phonon dispersion curves\cite{Stassis} are listed for comparison.
The theoretical values lie systematically 20-30 \% above the experimental values, which partly is due to the too low equilibrium volume obtained in the LDA. The
fact that the experiments are done at room temperature, while the calculations pertain to zero temperature also contribute to this discrepancy.
To the best of our knowledge no experimental determinations of the elastic constants of YSn$_3$ have been reported. Neither have any theoretical determinations been reported.
From the calculated values of the elastic constants, it can be seen that they satisfy the mechanical
  stability criteria\cite{Huang} for a cubic crystal i.e.  $C_{11}> C_{12}$, $C_{44}> 0$, and $C_{11}  + 2C_{12}> 0$, consistent with the elastic stability of these materials. 
Pugh \cite{Pugh} has proposed a simple
relationship in which the ductile/brittle properties of materials could be related empirically to their elastic constants by the ratio $G_H/B$. 
If $G_H/B<0.57$, the materials behave in a ductile manner, and brittle otherwise. From the values of $G_H/B$  reported in Table II it emerges
that both compounds are of ductile character, and that
YSn$_3$ is more ductile than LaSn$_3$. Cauchy's pressure ( $C_p=C_{12}-C_{44}$ ) is  another index to determine the ductile/brittle nature of metallic
compounds, where a  positive value of Cauchy's pressure indicates ductile nature, while a negative value indicates a brittle nature of the compounds. 
The calculated  positive values of the Cauchy's pressure reported in Table II corroborate the ductile nature of LaSn$_3$ and YSn$_3$. 
The Young's modulus $E$ also
reflects the ductility. The larger the value of $E$, the stiffer is the material, and as the covalent nature of the compounds increases $E$ also 
increases. Another important parameter describing the ductile nature of solids is the
Poisson's ratio $\sigma$ (see Table II), which is calculated using the formula given in Ref. \onlinecite{kanchana}. For ductile metallic materials $\sigma$ is
 typically around 0.33.\cite{Haines} So the ductility of these compounds is confirmed by the calculated values of $\sigma$ reported in Table II. The 
anisotropy $A$ is defined as the ratio between $C_{44}$ and $(C_{11} - C_{12})/2$, which becomes unity for an isotropic system. According to this
definition, LaSn$_3$ and YSn$_3$ are elastically anisotropic. 

The elastic constants and bulk modulus increase monotonously under compression, fulfilling the mechanical stability criteria also at higher pressures. 
Having calculated  Young's modulus E, the bulk
modulus B, and the shear modulus G, one may derive the Debye temperature  using Eq. (1). The calculated sound velocities ( $v_l$, $v_t$ and
$v_m$ ) and the Debye temperature ( $\Theta_D$) are included in Table II. The experimental Debye temperatures are almost the same for LaSn$_3$ and YSn$_3$, while the present calculations
find the Debye temperature of YSn$_3$ substantially lower than that of LaSn$_3$, a consequence of the smaller calculated values of the elastic constants of YSn$_3$.   The sound
velocities  increase with  pressure for LaSn$_3$ and YSn$_3$, reflecting the increase
in the relevant acoustic phonon frequencies with pressure.

\subsection{Band structure and Density of States under pressure}

 The  band structures of LaSn$_3$ and YSn$_3$ are illustrated in Fig. 1. The electronic levels are calculated along the high-symmetry directions with and without spin-orbit coupling and
 at the LDA equilibrium volumes. The band structure of LaSn$_3$ with SOC
 compares well with  earlier work. \cite{Yamagami,Boulet}
 Overall, the  band structures of LaSn$_3$ and YSn$_3$  are very similar, as noted by Ref. \onlinecite{Dugdale}, which also discussed the effect of SOC.   
 The major difference between the two compounds in the vicinity of the Fermi level, E$_F$, occurs at the X-point, where a band crosses the $E_F$ for LaSn$_3$, but stays below $E_F$ for YSn$_3$. This gives rise to a small hole pocket around the $X$-point in LaSn$_3$, which does not appear for YSn$_3$ (and which would not be 
if SOC is not included).
\textcolor{blue}{ \bf A second, less significant feature, is a very dispersive band along $\Gamma-R$, which dips below 
 $E_F$ for YSn$_3$, but stays above $E_F$ for LaSn$_3$. This is illustrated in the inset of Fig. 1.} This band contributes to the third Fermi surface of YSn$_3$, which  in LaSn$_3$ only appears under pressure.
 
 The main interesting pressure effect on the  electronic structure of these compounds is the opposite movement of the valence band and the conduction band.
  Along all high symmetry directions 
  the valence bands move
 down, while an upward shift of the conduction bands under compression is seen. This is illustrated for LaSn$_3$ in Fig. 2, which shows the band structure for $V/V_0=1.0$ and $V/V_0=0.90$, where $V_0$ denotes the experimental equilibrium volume. Owing to this opposite movement of bands under pressure the number of electron states in the pockets around the M-points, as well as the number of hole states around the X-points increase.
 
 The narrow bands around 1.5-2 eV above the Fermi level (see Fig. 1(a)) in LaSn$_3$ are the La - $4f$ bands. These unocupied La - $4f$ bands overlap with the La - $5d$ bands, 
 with some influence on the energy band structure in the vicinity of the Fermi energy. This is illustrated with the density of states, which is shown in Fig. 3, for both LaSn$_3$ and YSn$_3$.
 In both compounds, the dominating character around the Fermi level is from Sn $5p$, with appreciable admixture of Y $4d$ or La $5d$. But for LaSn$_3$ the tail of the $4f$ (blue line) in Fig. 3(a) also crosses the Fermi level. The behaviour of the total density of states at the Fermi level
 under compression is shown in Fig. 4. While the  total density of states at the Fermi energy, $N(E_F)$, decreases smoothly for YSn$_3$ under compression, it is more irregular for LaSn$_3$. 
   In LaSn$_3$, $N(E_F)$ passes through a minimum at $V/V_0=0.94$ (pressure of 1 GPa according to LDA). This behaviour of  LaSn$_3$ is caused by the occurrence of the third Fermi sheet under pressure, 
 which is discussed in the next subsection. 
 
 In Table III the density of states at the Fermi level is compared to the experimental value as derived from the Sommerfeld coefficient. From the difference, the average electron-phonon coupling constant
 $\lambda_{ep}$ may be estimated, assuming
\begin{equation}
	\frac{\gamma^{expt}}{\gamma^{calc}}=1+\lambda_{ep}.
\end{equation}
For LaSn$_3$ this reaches a value of $\lambda_{ep}=0.86$, while for YSn$_3$  a value of $\lambda_{ep}=0.34$ is found. Note that the latter is lower than the value deduced by Dugdale, Ref. \onlinecite{Dugdale}, because the inclusion of spin-orbit coupling in the calculation increases $N(E_F)$ by $\sim 20$ \%. The value obtained here for LaSn$_3$ is in good agreement with the value
found by Ref. \onlinecite{Toxen2}.
The superconducting transition temperature for LaSn$_3$ may be estimated using the McMillan formula\cite{McMillan}:
\begin{equation}
	T_c=\frac{\Theta_D}{1.45}\exp\left(-\frac{1.04(1+\lambda_{ep})}{\lambda_{ep}-\mu^*(1+0.62\lambda_{ep})}   \right).
\end{equation}
Using a typical value of $\mu^*=0.12$ (Ref. \onlinecite{Dugdale}) and the Debye temperature of LaSn$_3$ of $\Theta_D=205$ K, this leads to a calculated value of $T_c=8.1$ K, which is in excellent agreement 
with the experimental value of $T_c=6.5$ K, given the 
uncertainty of the parameters of the McMillan formula.

\subsection{Fermi surface under pressure}

 The Fermi surfaces of LaSn$_3$ and YSn$_3$ at the experimental equilibrium volumes are shown in Fig. 5. The similarity of the band structures lead to nearly identical topology of 
 the Fermi surfaces for the two compounds. The major difference is that for LaSn$_3$ two bands
 cross the Fermi level, whereas for YSn$_3$ three bands cross the Fermi level. The first Fermi surface sheet, Fig. 5(a) and 5(c), is a hole pocket 
 centered on the $\Gamma$ point. The second sheet, Fig. 5(b) and 5(d), is a very complex surface, which we discuss in detail for LaSn$_3$ below. 
 In comparison between the two compounds, a small hole pocket is seen around the X-point in Fig. 5(b), which is absent in Fig. 5(d), which reflects the difference in band structure at $X$ as discussed in the previous section. Finally, the third surface of YSn$_3$ is a small electron part close to the X-point.
In Fig. 6 we illustrate for LaSn$_3$ the complexity of the second sheet by showing horizontal cuts through the three-dimensional Fermi surface.  
It is to be noted that without taking into account the SOC three bands would cross the
Fermi level even at ambient volume, and the corresponding Fermi surfaces would be similar for 
LaSn$_3$ and YSn$_3$, and furthermore, for YSn$_3$ the Fermi surface would be topologically the  same with and without SOC.  

 The most striking change in the Fermi surface
of LaSn$_3$ under pressure is the appearance of the third surface, already seen in YSn$_3$ at ambient conditions. This appears at a compression of
$V/V_0=0.94$, as shown in Fig. 7(c).
A second, less drastic change occurs in the second Fermi surface of LaSn$_3$, where a 
change in topology is observed. This is most easily seen in the two-dimensional contours of Fig. 6(b), where the hole pocket around the X-point, (the middle point of the $k_z=\pi/a$ face)
increases and merges with the surrounding triangular hole regions. In contrast, for $V=V_0$, Fig. 6(a), this pocket is detached from the larger hole region, facilitating a small closed electron orbit. 
At the same time, the electron concentration around the M-point (the midpoints of all edges of the BZ) increases under compression, which eventually leads to the connection of all electron pockets in the $k_z=\pi/a$ face. This
happens around $V/V_0=0.90$ (see Fig. 7(b)). 
Altogether, under compression the electron concentration at M and the hole concentration at X
increase simultaneously in LaSn$_3$.
In the case of YSn$_3$ only the electron concentration at the M-point increases, while there is no hole pocket at X even at ambient volume, and therefore the Fermi surface topology of YSn$_3$
remains unchanged under (modest) compression.

\par

The occurrence of a third Fermi surface sheet for LaSn$_3$ under pressure leads to an increase of the density of states at the Fermi level, as illustrated in Fig. 4.
For comparison the same quantity for YSn$_3$ is also shown in the figure, and it is seen to decrease monotonously under pressure.  
These results for LaSn$_3$ are in accordance with the 
zero-pressure  measurements of the superconducting transition temperature in the (La,Th)Sn$_3$ alloy system as investigated by Havinga et al.\cite{Havinga} 
These authors also speculated that their observed oscillatory behaviour of $T_c$ versus alloy composition might be due to a
singular behaviour of the electronic
density of states in the vicinity of the Fermi level of LaSn$_3$. 
Huang et al. \cite{HuangSSC}
observed an irregular behaviour of T$_c$ for LaSn$_3$ under pressure.
These authors reported an initial slight increase in T$_c$ with a maximum at a pressure around 0.8 GPa, beyond which T$_c$ gradually decreases. 
Within the BCS framework of superconductivity, the change in T$_c$ observed could reflect a change in
 the density of states at the Fermi level. At first, this seems in accordance with the non-monotonic variation under compression
 of the density of states at the Fermi level 
in LaSn$_3$ found in the present calculations. However, we find the opposite trend of an initially decreasing density of states at the Fermi level, and an increase only starts for compressions below 0.94$V_0$. Several other factors of course also influence the transition 
temperature. The pressure dependence of T$_c$ arising from pressure-induced abrupt changes in the Fermi surface topology
was theoretically analysed by Makarov and Baryakhtar \cite{Makarov}.

		
\section {Conclusion}

An {\it ab initio} study of the intermetallic compounds LaSn$_3$ and YSn$_3$ was performed within the local density approximation. 
The structural, electronic, elastic, and mechanical properties as well as the Fermi surfaces were studied, including  pressure effects. These compounds are ductile in nature and their Cu$_3$Au crystal structure is stable even at high pressure. 
The elastic constants and the bulk modulus increase
monotonically with pressure.  The density of states near to the Fermi level are mainly  Sn p-like states in both compounds, but a
Fermi surface topology change is observed in LaSn$_3$ at a compression of around $V/V_0\sim 0.94$, where 
a third Fermi sheet occurs. 
     	
\clearpage
*Author for Correspondence, 
E-mail: kanchana@iith.ac.in

\newpage

\begin{table}[th]
\begin{tabular}{ccccc}
\hline
Parameters             &        & LaSn$_3$                      &     &        YSn$_3$      \\
\hline
                  $a$  & GGA    & 4.81                          &     &         4.73        \\
                       & LDA    & 4.70                          &     &         4.61        \\
                       & Theory$^a$  & 4.73                     &     &         -             \\
                       & Expt.  & 4.774$^b$,4.769$^c$           &     &        4.667$^d$    \\                       
                  $B$  &  GGA   &   55.5                        &     &         56.9         \\
                       & LDA    &   68.2                        &     &        70.6         \\
                       & Theory$^a$ & 78                        &     &        -              \\
                       & Expt.  &   51.5$^e$                    &     &        -           \\

\hline
\end{tabular}
\caption{Calculated lattice constant $a$ (in \AA) and bulk modulus $B$ (in GPa)  for LaSn$_3$ and YSn$_3$, as obtained with the
GGA and LDA approximations for exchange and correlation. The bulk modulus is evaluated at the theoretical equilibrium volumes.
Experimental values are quoted for comparison 
a: Ref. \onlinecite{Shao-ping}; 
b: Ref. \onlinecite{Gambino}; 
c: Ref. \onlinecite{Landelli}
d: Ref. \onlinecite{Kawashima};
e: From elastic constants obtained in Ref. \onlinecite{Stassis}}
\end{table}

\newpage
\begin{table}[h]
\begin{tabular}{ccc}
\hline
Parameters             &    LaSn$_3$    &    YSn$_3$  \\
\hline
   C$_{11}$ (GPa)      &     97.3 (70.5$^a$)       &    82.3     \\
   C$_{12}$ (GPa)      &     53.6 (42.0$^a$)       &    64.6     \\
   C$_{44}$ (GPa)      &     44.2 (33.5$^a$)       &    34.2     \\
       $A$             &     2.02                  &    3.85     \\
$G_H$ (GPa)            &     33.3                  &    20.0     \\
$E$ (GPa)              &     85.9 (64$^b$)         &    54.9 (98$^b$)     \\
$\sigma$               &     0.29                  &    0.37     \\
$G_H/B$                &     0.49                  &    0.28     \\
$C_p$ (GPa)            &      9.4                  &    30.3     \\
$v_l$ (km/s)           &      3.77                 &    3.59     \\
$v_t$ (km/s)           &      2.05                 &    1.63     \\
$\Theta_D$ (K)         &      230 (205$^c$)        &   188.4 (210$^d$)    \\

\hline
\end{tabular}
\caption{Elastic constants and derived quantites for LaSn$_3$ and YSn$_3$, as calculated with LDA at equilibrium volume. 
$A$ is the anisotropy factor, $A=2C_{44}/(C_{11} - C_{12})$, and $C_p=C_{12}-C_{44}$ is the Cauchy pressure. 
Experimental values are given in parentheses where available.
a: From phonon measurements, Ref. \onlinecite{Stassis};
b: Ref. \onlinecite{Dudek}
c: Ref. \onlinecite{Bucher};
d: Ref. \onlinecite{Kawashima}.
}
\end{table}

\newpage
\begin{table}[th]
\begin{tabular}{llcccc}
\hline
             &            & DOS(states/eV)        & $\gamma$ (mJ/mol K$^2$)&     $\lambda_{ep} $   &   $T_c$ (K)    \\
\hline
    LaSn$_3$ & Theory$^a$ &       2.67            &         6.28           &         0.86          &    8.1         \\
             & Theory$^b$ &       2.15            &          6.03                    &  0.82                 &   -    \\
             & Expt.      &       2.6$^c,$2.8$^d$ &         11.66$^d$,10.96$^e$,11.0$^f$      &         0.8$^c$       &    6.45$^g$,6.02$^h$     \\

     YSn$_3$ & Theory$^a$ &       2.41            &         5.67           &         0.34          &    0.11        \\
             & Theory$^i$ &       1.92            &         4.53           &         0.63,0.99     &    5.93        \\
             & Expt.      &       -               &         7.57$^j$       &         -             &    7.0$^j$     \\

\hline
\end{tabular}
\caption{Calculated density of states at the Fermi level (evaluated at the experimental equilibrium volumes), together with derived
Sommerfeld constants, $\gamma$, and electron-phonon coupling constants, $\lambda_{ep} $,  for LaSn$_3$ and YSn$_3$.  Last column gives the superconducting transition temperature.
Experimental values are quoted for comparison.
a: This work, LDA; 
b: Ref. \onlinecite{Shao-ping}
c: Ref.  \onlinecite{Toxen2}; 
d: Ref.  \onlinecite{Maple};
e: Ref.  \onlinecite{HuangSSC};
f: Ref.  \onlinecite{Stassis};
g: Ref. \onlinecite{Gambino}; 
h: Ref. \onlinecite{Havinga};
i: Ref. \onlinecite{Dugdale};
j: Ref. \onlinecite{Kawashima};
}
\end{table}

\newpage

\begin{figure}

\subfigure[]{\includegraphics[width=80mm,height=90mm]{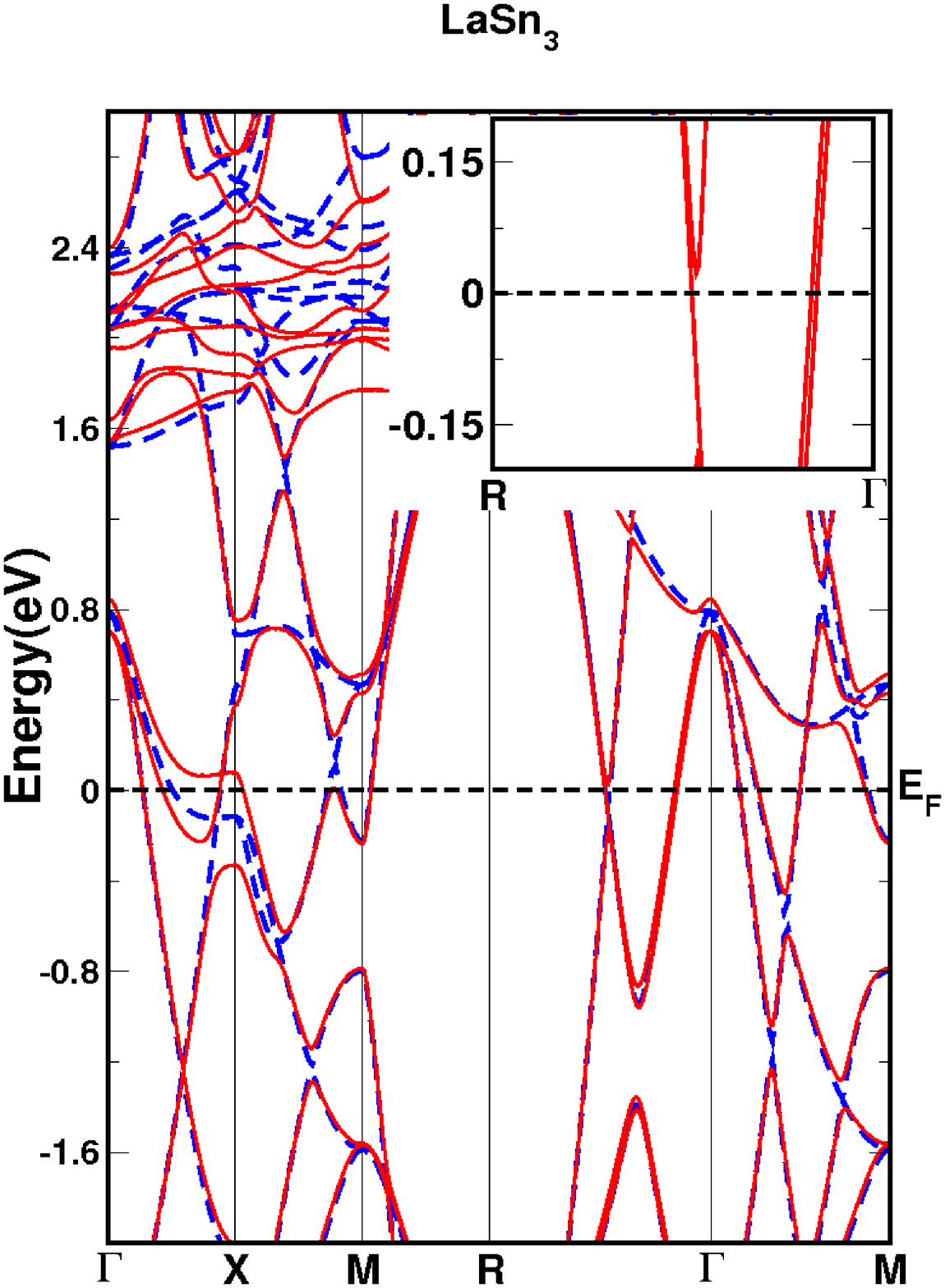}} 
\subfigure[]{\includegraphics[width=80mm,height=90mm]{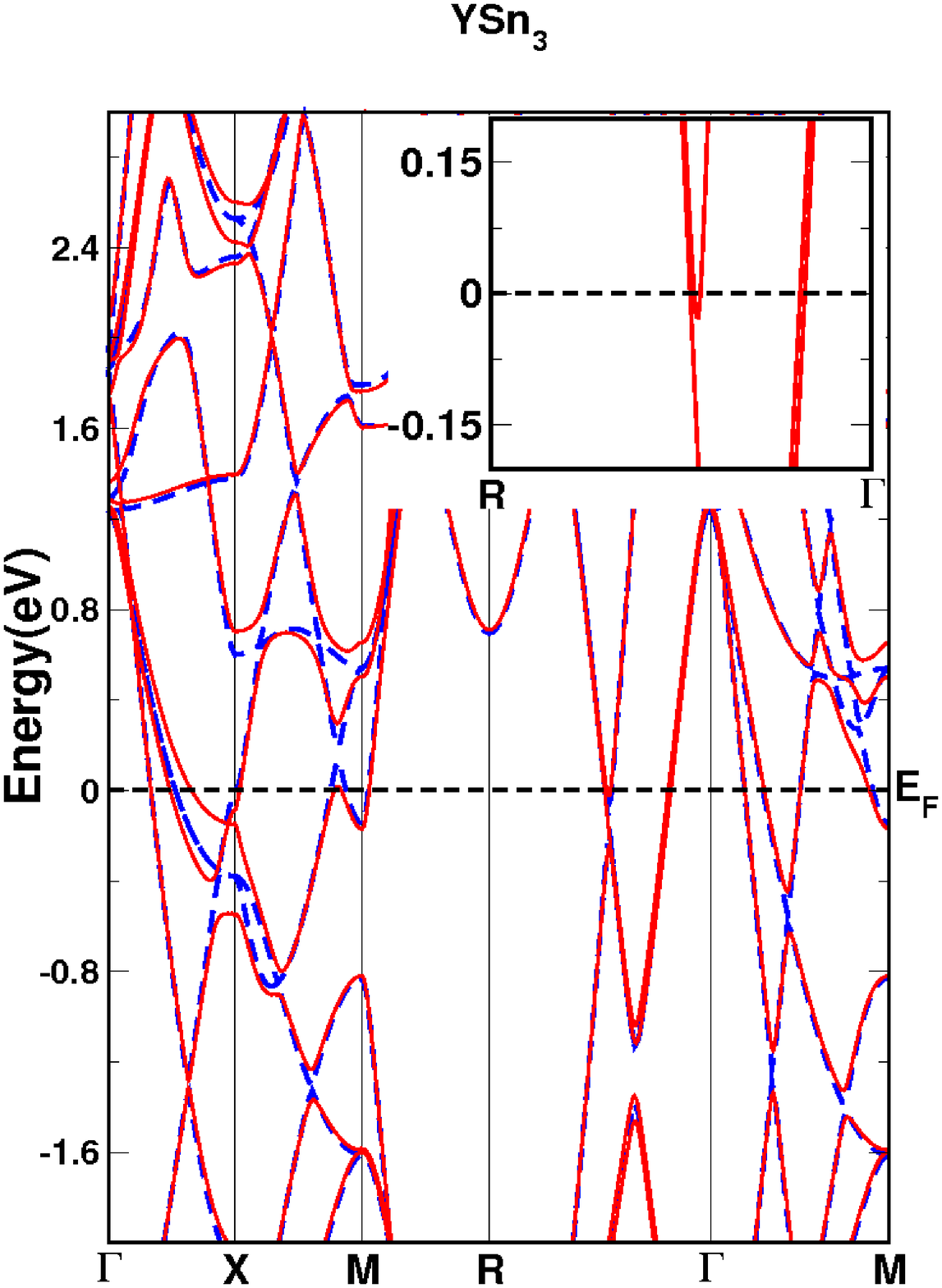}} 

\caption{(Color online) Electronic band structures of (a): LaSn$_3$ and (b): YSn$_3$. The solid lines
 (red colour) show the electronic levels calculated with spin-orbit
coupling included, while the dotted lines (blue colour) show the electronic levels as calculated without  spin-orbit coupling. The energies are given
in eV relative to the Fermi level, $E_F$, which is marked with the horizontal dashed line.
The major difference between the two compounds  around the Fermi level occurs in the vicinity of the X-point (for SOC included). \textcolor{blue}{ \bf A second, less significant feature, is a very dispersive band along $\Gamma-R$, which 
stays above $E_F$ for LaSn$_3$, but dips below $E_F$ for YSn$_3$. The inset illustrates this.}}
\end{figure}

\begin{figure}
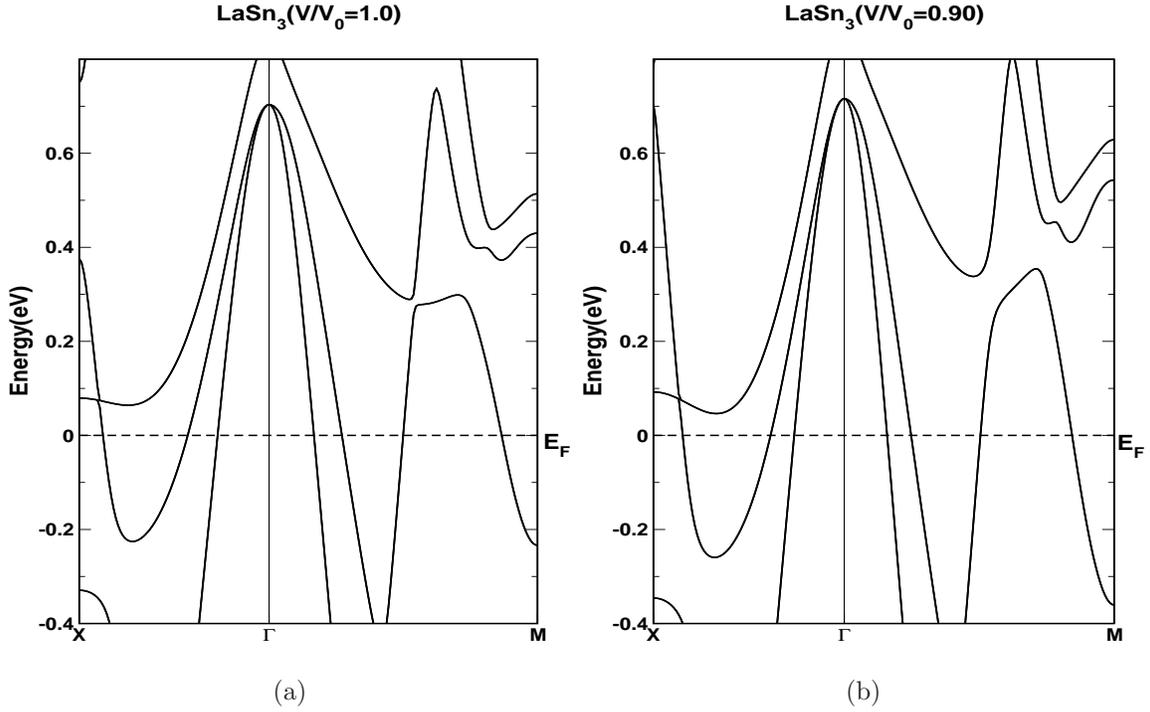

\subfigure[]{\includegraphics[width=75mm,height=85mm]{Fig-3.eps}}
\subfigure[]{\includegraphics[width=75mm,height=85mm]{Fig-4.eps}}
\caption{Band structure of LaSn$_3$ under compression (zoom-in on the vicinity of the Fermi level). The electron pocket at $M$ and the hole pocket at $X$ increases under pressure.
} 

\end{figure}

\begin{figure}
\subfigure[]{\includegraphics[width=65mm,height=60mm]{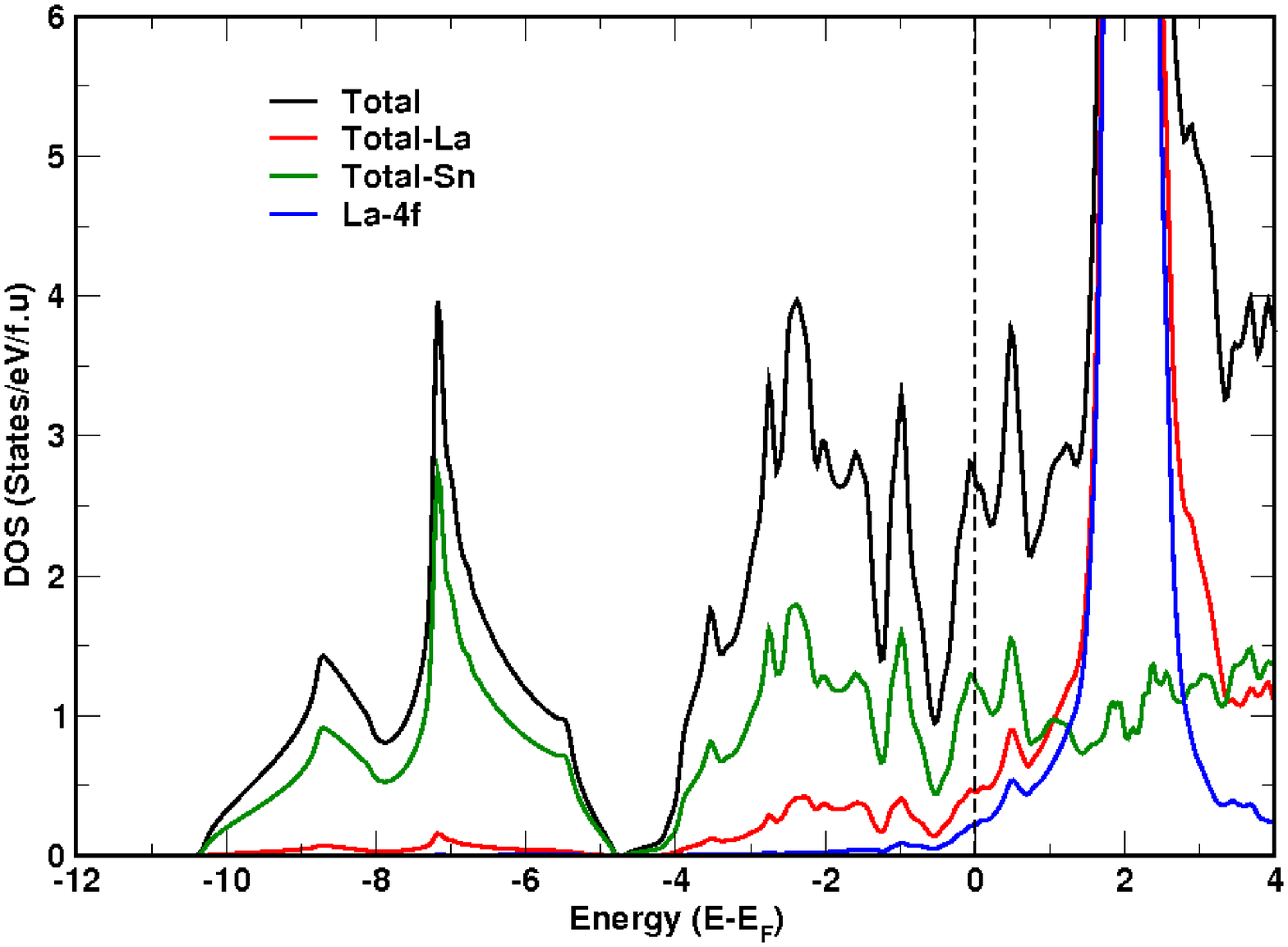}}\vspace{0.2mm}
\subfigure[]{\includegraphics[width=65mm,height=60mm]{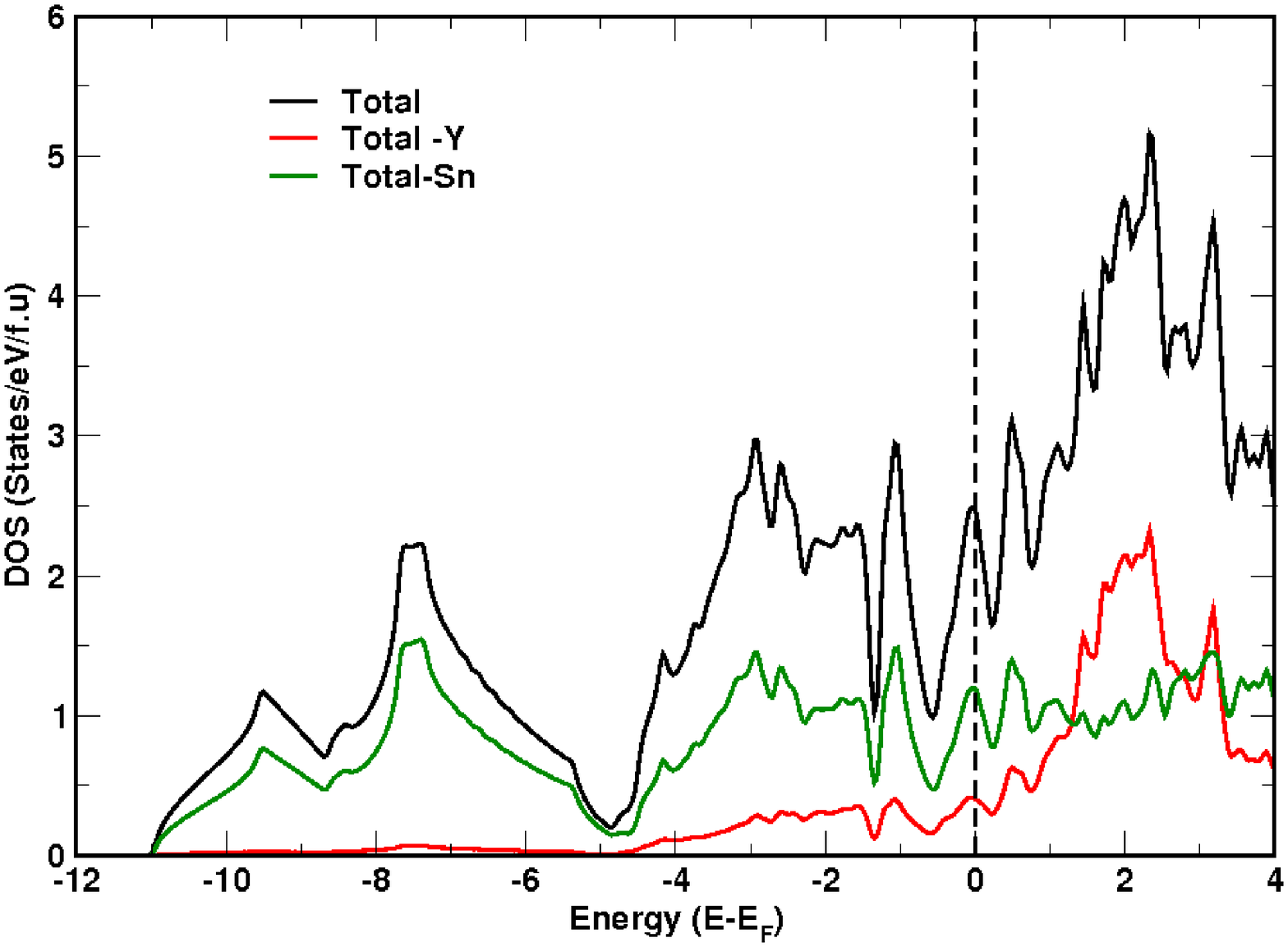}}
\caption{(Color online) Density of states of (a): LaSn$_3$ and (b): YSn$_3$, as calculated at the experimental lattice constants.
The total DOS as well as La, Sn and La-$4f$ partial contributions are shown in (a),
the total DOS as well as Y and Sn partial contributions are shown in (b). The unit is states per eV and per formula unit. A factor of 2 for spin is included.}
\end{figure}

\begin{figure}
\begin{center}
\subfigure[]{\includegraphics[width=95mm,height=70mm]{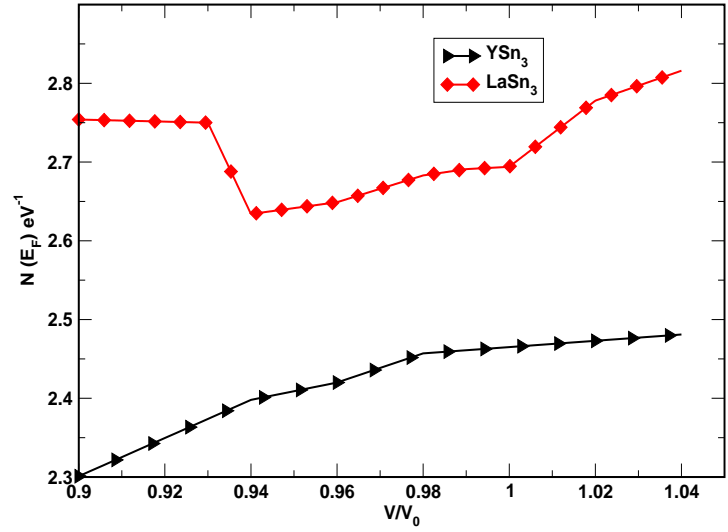}}
\caption{(Color online) Density of states at the Fermi level, $N(E_F)$, for LaSn$_3$  and YSn$_3$  under compression. The jump in density of states for LaSn$_3$ 
around $V/V_0=0.94$ is due to the appearance of the third Fermi sheet. $V_0$ denotes the respective experimental equilibrium volumes of
LaSn$_3$  and YSn$_3$.}
\end{center}
\end{figure}

\begin{figure}
\subfigure[]{\includegraphics[width=64mm,height=64mm,angle=0]{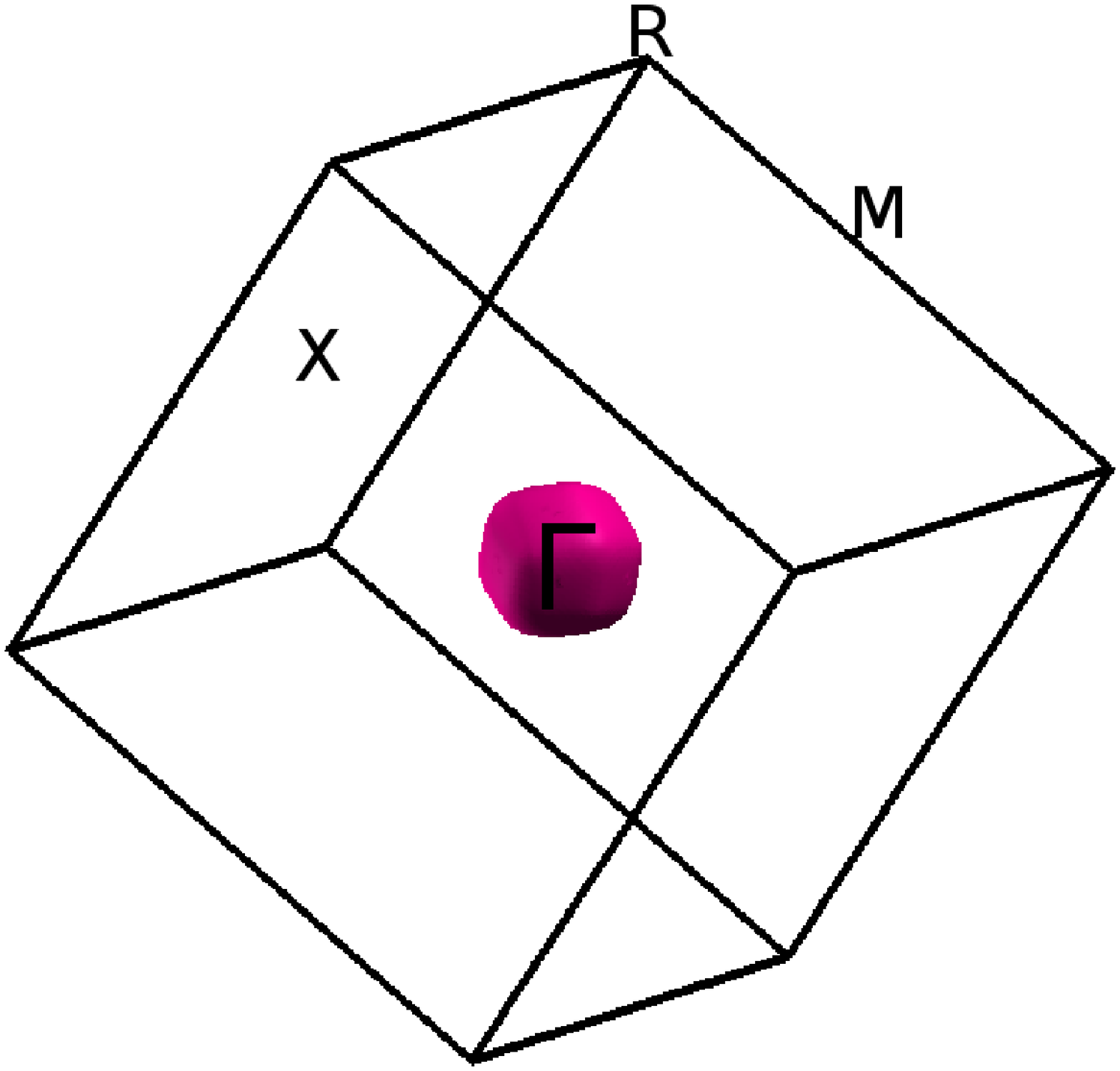}}\hspace{0mm}
\subfigure[]{\includegraphics[width=64mm,height=64mm,angle=0]{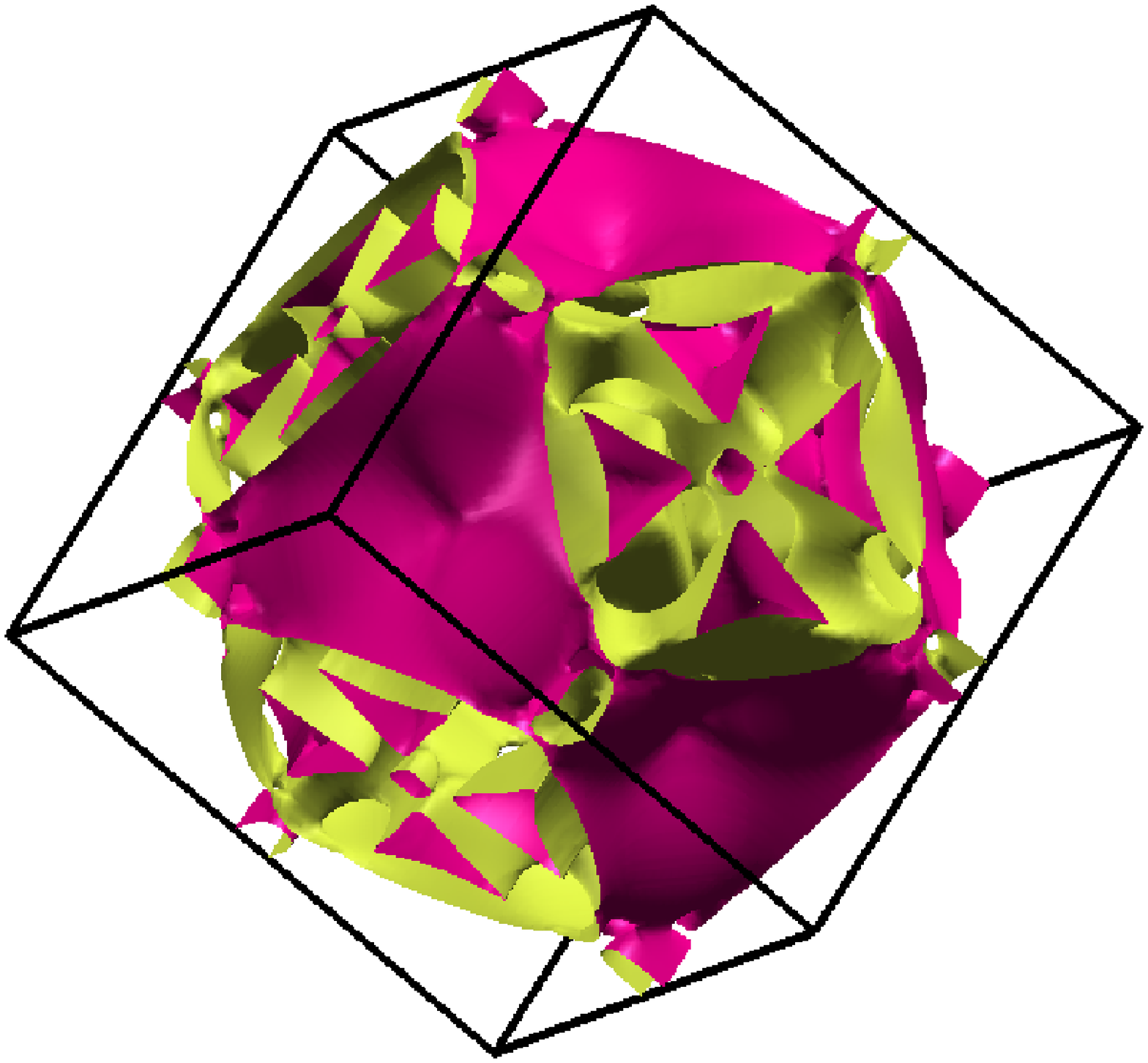}}\\
\subfigure[]{\includegraphics[width=64mm,height=64mm]{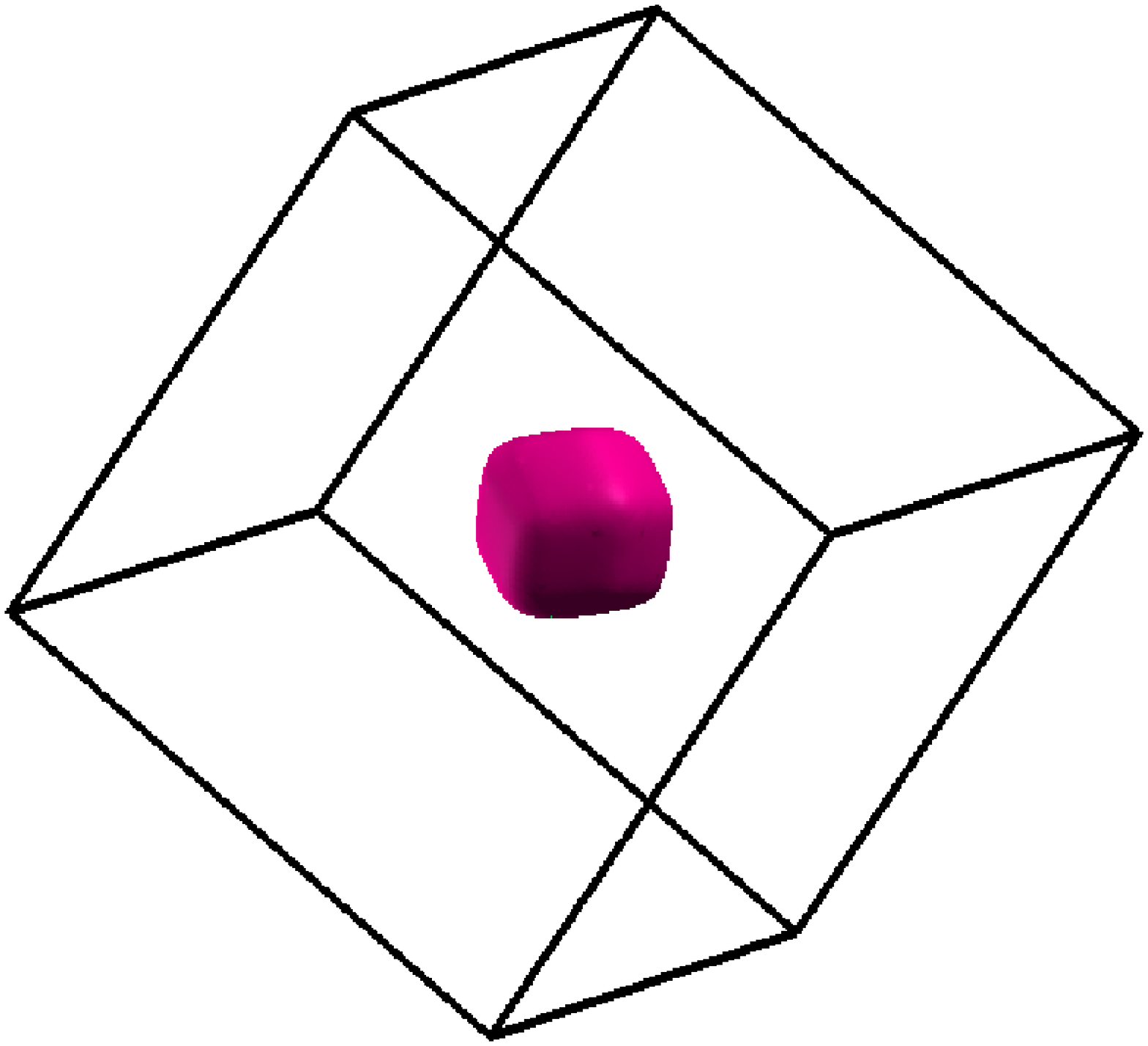}}\hspace{0mm}
\subfigure[]{\includegraphics[width=64mm,height=64mm]{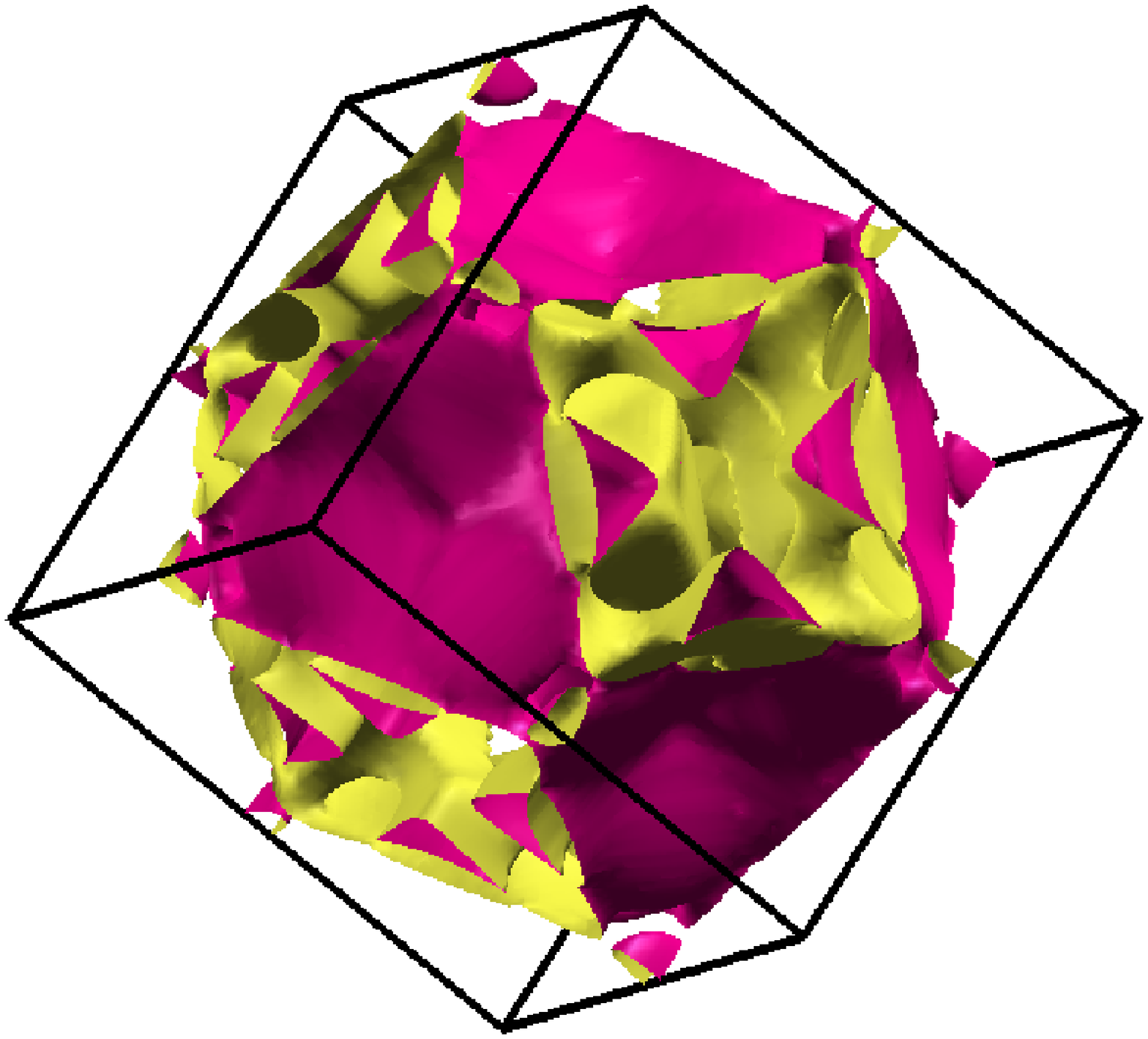}}\hspace{0mm}
\subfigure[]{\includegraphics[width=64mm,height=64mm]{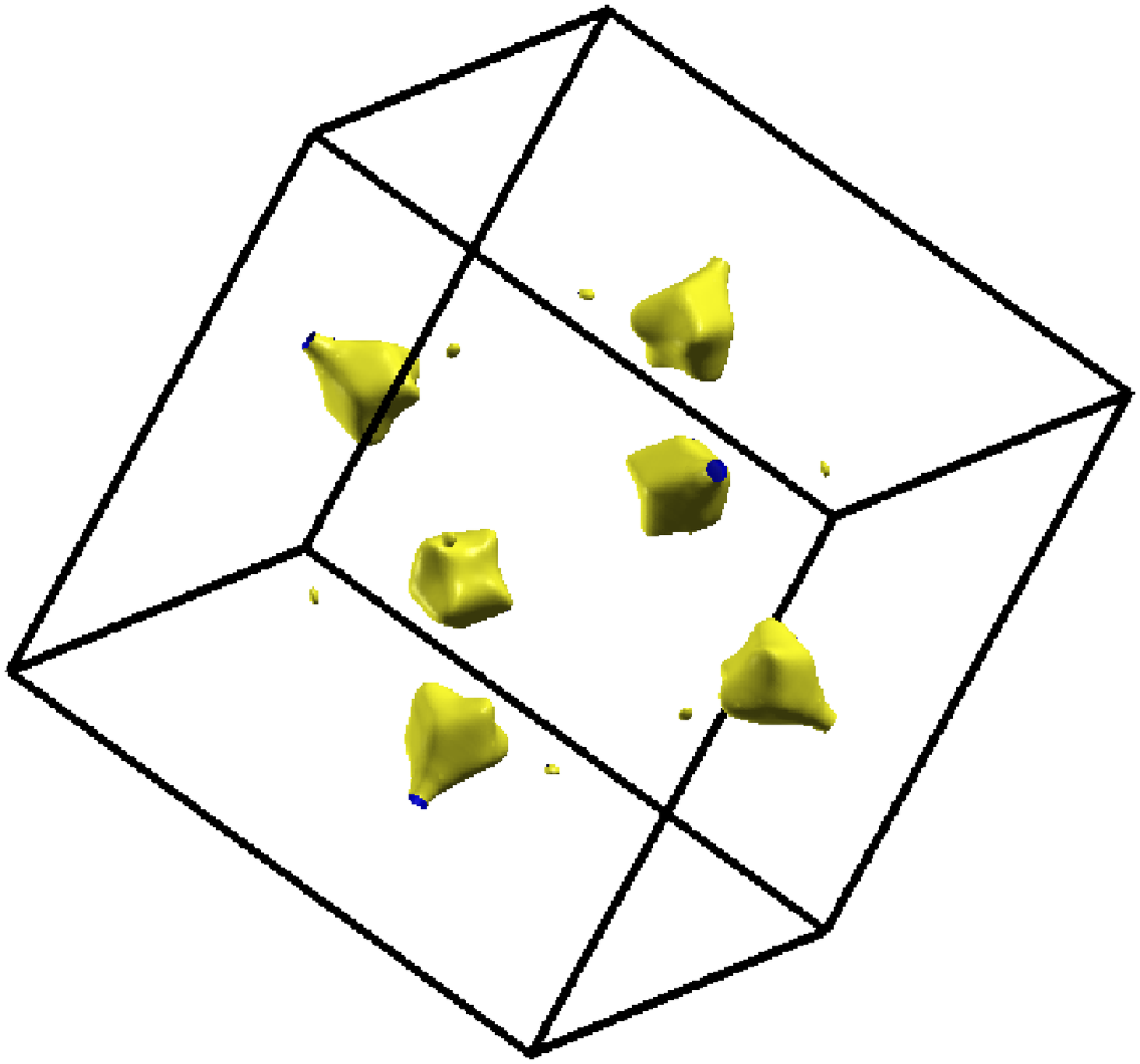}}
\caption{(Color online) Fermi surface of (a), (b): LaSn$_3$.  and (c), (d), (e): YSn$_3$ (including spin-orbit coupling and evaluated at the experimental equilibrium volumes).
(a) and (c) are hole pockets around $\Gamma$, (e) are electron pockets around the X-points. 
The complex second sheet of (b) is illustrated through two-dimensional cuts
in Fig. 6(a). In (a) the BZ critical points are marked. }
\end{figure}
\newpage
\begin{figure}[h]
\includegraphics[width=100mm,height=60mm]{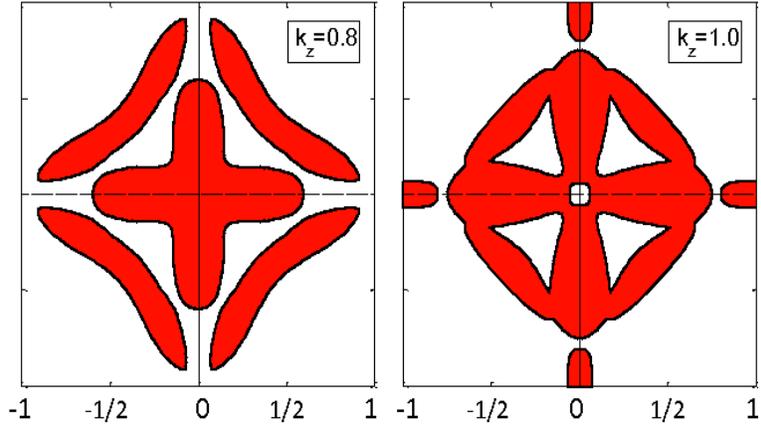}
\vspace{-5mm}
\begin{center}(a) \\[5mm] \end{center}
\includegraphics[width=100mm,height=55mm]{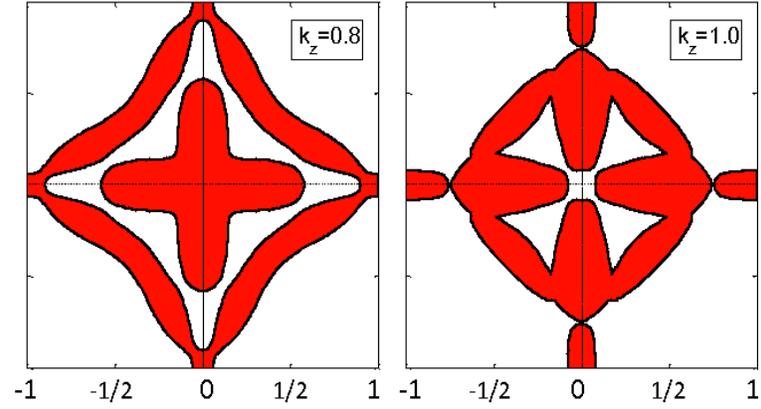}
\vspace{-5mm}
\begin{center}(b)\end{center}
\caption{(Color Online) Second Fermi surface of LaSn$_3$, 2-dimensional contours corresponding to $k_z=0.8$ and $1.0$ in units of $\pi/a$. 
(a): at the experimental equilibrium volume, and (b): at a volume of $90$ \% of the experimental equilibrium volume.
The shaded (red) areas correspond to
occupied states.}
\end{figure}

\begin{figure}
\subfigure[]{\includegraphics[width=64mm,height=64mm,angle=00]{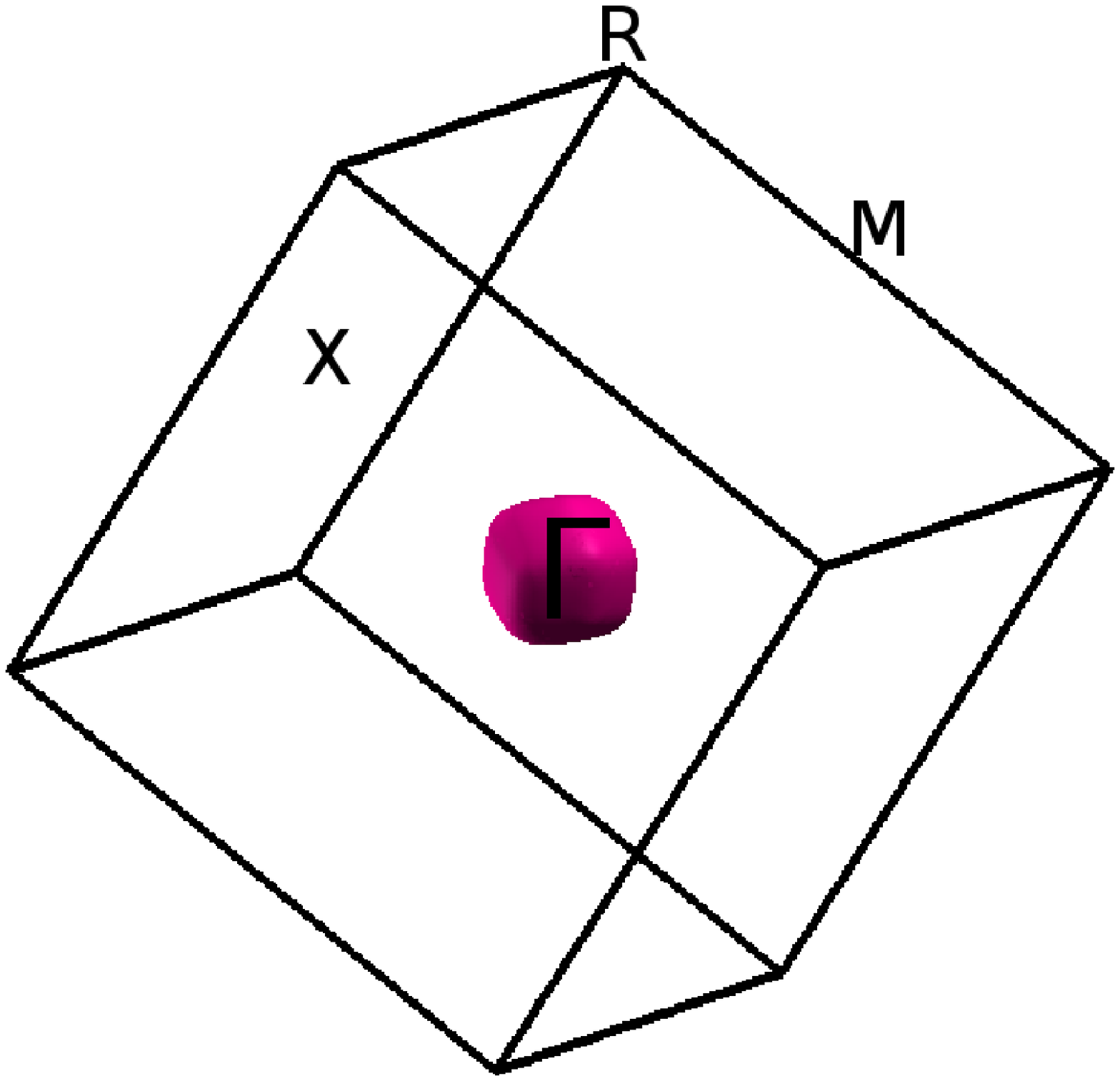}}\hspace{0mm}
\subfigure[]{\includegraphics[width=64mm,height=64mm]{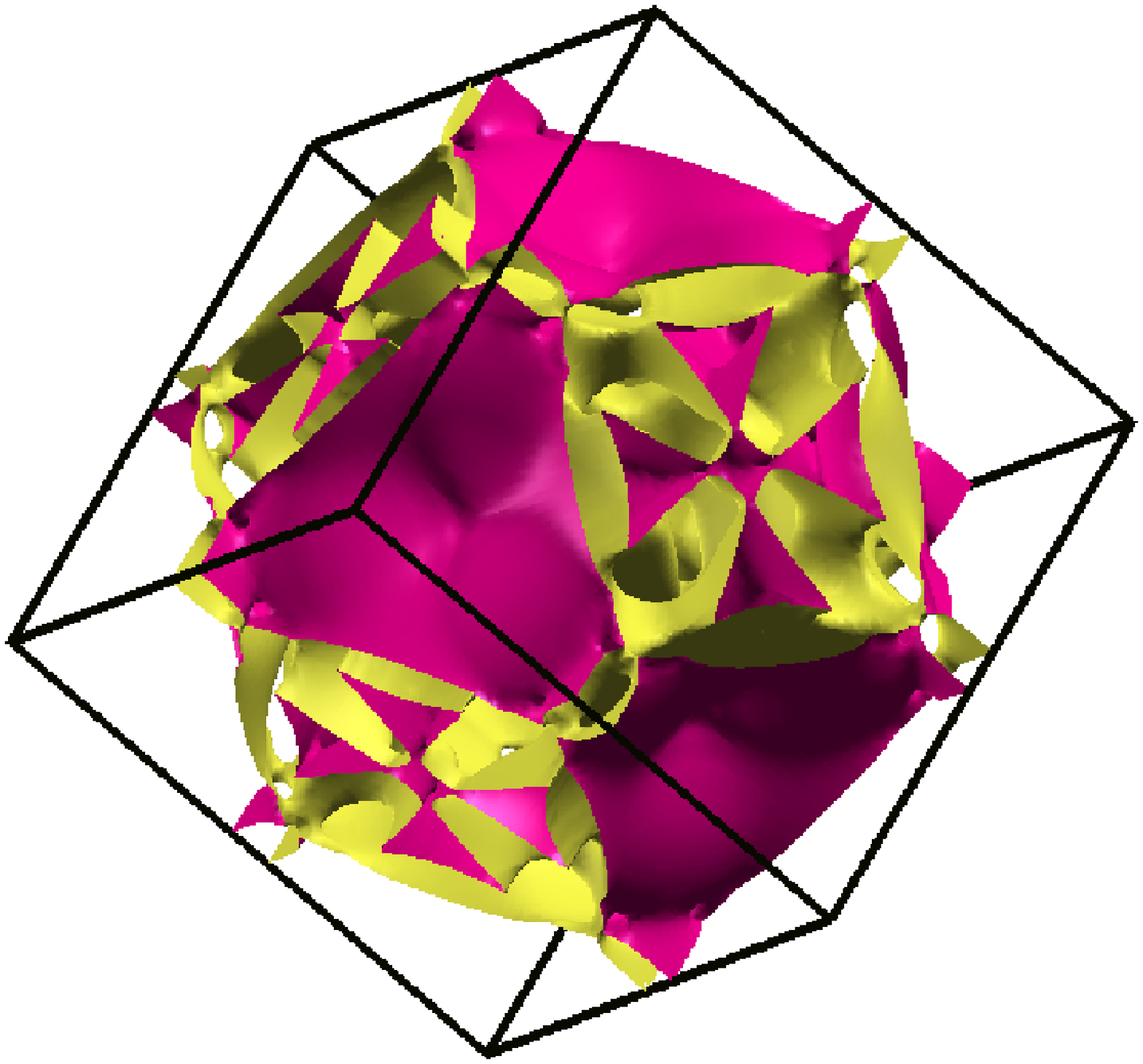}}\hspace{0mm}
\subfigure[]{\includegraphics[width=64mm,height=64mm,]{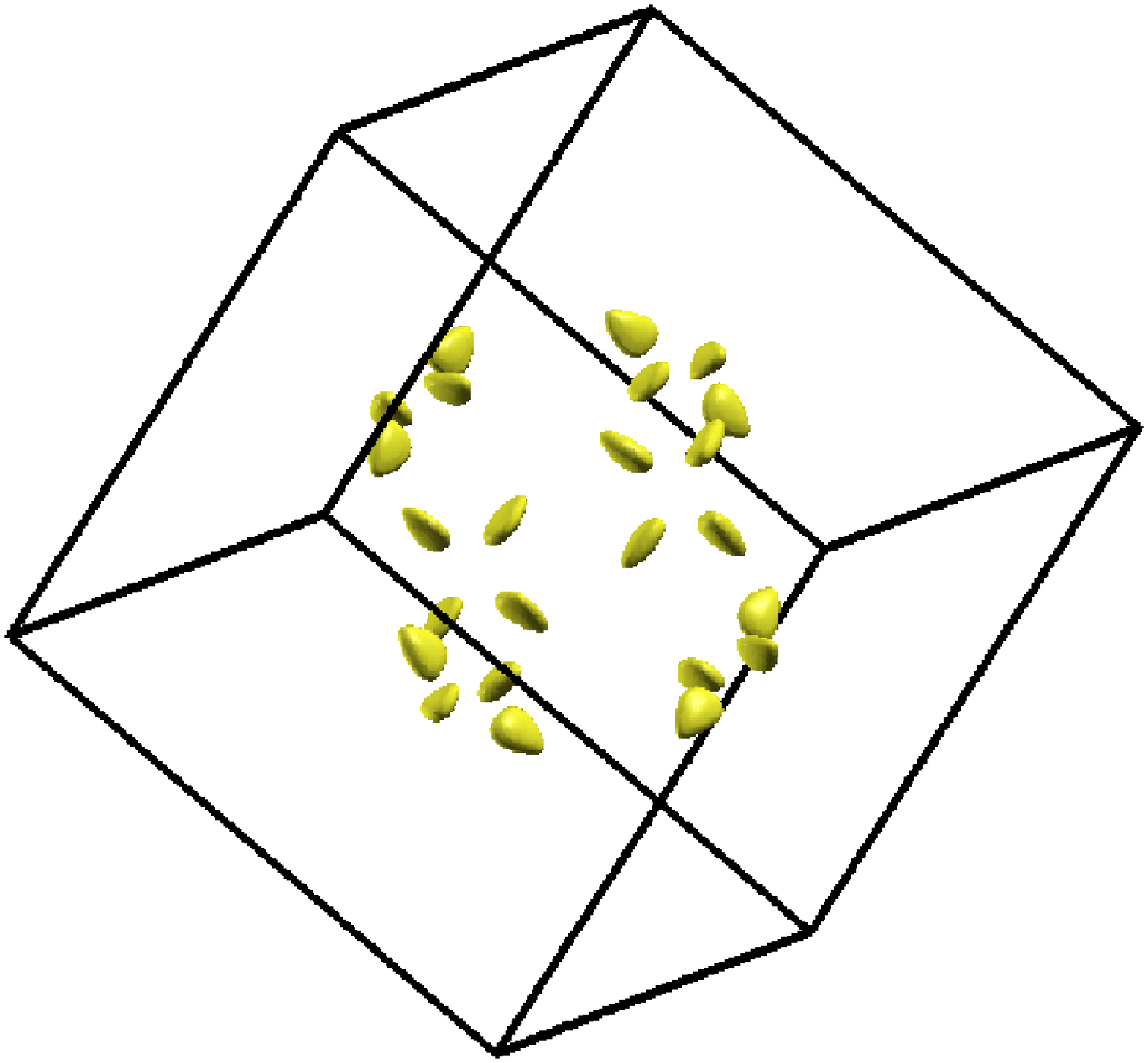}}
\caption{(Color online) Fermi surface of LaSn$_3$ at compression $V/V_0=0.90$. V$_0$ denotes the experimental equilibrium volume.
 }
\end{figure}

\end{document}